\documentclass[aps,prl,superscriptaddress,showpacs,twocolumn]{revtex4}%
\usepackage{amsmath}
\usepackage{graphicx}
\usepackage{amsfonts}
\usepackage{amssymb}%
\begin{document}
\title{Charged Snowball in Non-Polar Liquid}
\author{Yu.Chikina}
\affiliation{DRECAM/SCM/LIONS CEA - Saclay, 91191 Gif-sur-Yvette Cedex France}
\author{V.Shikin}
\affiliation{ISSP, RAS, Chernogolovka, Moscow District, 142432
Russia}
\author{A.A.Varlamov}
\affiliation{INFM-CNR, COHERENTIA, via del Politecnico, 1, "I-00133, Rome, Italy}
\date{\today}

\begin{abstract}
The problem of correct definition of the charge carrier effective
mass in superfluid helium is revised. \ It is demonstrated that
the effective mass $M$ of a such quasiparticle can be introduced
without use\ of the Atkins's idea concerning the solidification of
liquid $He$ in the close vicinity of ion \cite{atkins}. The
two-liquid scenario of the ''snowball'' mass formation is
proposed. The normal fluid contribution to the total snowball
effective mass, the physical reasons of its singularity and the
way of corresponding regularization procedure are discussed.
Agreement of the theory with the available experimental data is
found in the wide range of temperatures.

\end{abstract}
\maketitle

The ion-dipole interaction between the inserted charged particle
and the induced electric dipoles of surrounded atoms is one of the
interesting phenomena which take place in a non-polar liquid. It
is this interaction which is generally responsible for the
different salvations phenomena there \cite{conway}, while in
non-polar cryogenics liquids (like $He$, $Ne$, $Ar$, etc.) it
leads to so-called  ''snowball effect''. The latter consists in
the formation of a non-uniformity $\delta\rho\left( r\right)  $ in
a density of a liquid around the inserted charged particle.

The ion-dipole interaction $U_{i-d}\left(  r\right)  $ between the inserted
charged particle and solvent atoms in the simplest form can be written as%

\begin{equation}
U_{i-d}\left(  r\right)  =-\frac{\alpha}{2}\frac{e^{2}}{r^{4}}, \label{poten}%
\end{equation}
where $\alpha$ is the polarization of a solvent atom and $r$ is the distance
to the charged particle. The presence of attraction potential (\ref{poten}) in
equilibrium has to be compensated by the growth of the solvent density
$\delta\rho\left(  r\right)  $\ in direction of the charged particle. The
latter can be estimated, using the requirement of the chemical potential
constancy:
\begin{align}
\delta\mu\left(  r\right)   &  =P_{s}\left(  r\right)  v_{s}-\frac{\alpha}%
{2}\frac{e^{2}}{r^{4}}=0,\qquad\nonumber\\
\delta\rho\left(  r\right)   &  =\left(  \frac{\partial\rho}{\partial
P}\right)  P_{s}\left(  r\right)  \equiv\frac{1}{s^{2}}P_{s}\left(  r\right)
\label{chempot}%
\end{align}
where $P_{s}\left(  r\right)  $ is the local pressure around the ion, $v_{s}%
$\ is the volume of individual solvent atom and $s$ is the sound velocity. The
distance at which $P_{s}\left(  r\right)  $ reaches the value of pressure of a
solvent solidification $P_{s}^{s}$%
\begin{equation}
R_{s}^4=\frac{\alpha}{2}\frac{e^{2}}{P_{s}^{s}v_{s}} \label{atkrad}%
\end{equation}
corresponds to the radius of rigid sphere which is called \textquotedblleft
snowball\textquotedblright. In the case of positive ions being in liquid
helium, where the inter-atomic distance $a\approx3$ $\mathring{A}$ and the
pressure $P_{s}^{s}\approx25$ $atm$, the snowball radius is estimated as
$R_{s}^{He}\approx7$ $\mathring{A}.$

Described above so-called Atkins's snowball model \cite{atkins} is
quite transparent and it was found useful for the various
qualitative predictions. In particular, it provides by the natural
definition and estimation for the effective mass $M$ of a such
quasi-particle as the sum of the extra mass caused by the presence
of the density perturbation $\delta\rho\left( r\right)  $
\begin{equation}
M^{\left(  sol\right)  }=4\pi\int_{a}^{\infty}r^{2}\delta\rho\left(  r\right)
dr\label{massa}%
\end{equation}
and associated mass, related to the appearance of the velocity
distribution around the sphere moving in an ideal liquid
\begin{equation}
M_{0}^{\left(  \mathrm{ass}\right)  }\left(  \rho,R_{s}\right)  =\frac{2}%
{3}\pi\rho R_{s}^{3}.\label{massb}%
\end{equation}
Both of these contributions turn out to be of the same order.
Calculated in this way snowball effective mass turns out to be
$M\geq 50m_{4},$ what roughly corresponds to the experimental data
\cite{Poitr}.

Careful analysis shows that both expressions (\ref{massa}) and
(\ref{massb}) require more precise definitions and further
development of the Atkins's model. Indeed, one can easily see that
the value of $M^{\left(  sol\right)  }$ turns out to be critically
sensitive to the lower limit of the integral. Atkins used the
value of helium inter-atomic distance $a\approx3$ $\mathring{A}$
\cite{atkins} as a rough cut-off parameter only. The microscopic
analysis shows (see Eq. (\ref{rid}) below) that the real lower
limit turns out to be less that this value.

Principal revision requires the definition of the associated mass
$M^{\left( \mathrm{ass}\right) }$ for the two-fluid model. The
matter of fact that its strong temperature dependence, which has
been systematically observed experimentally \cite{Dahm}
\cite{Mellor2}, cannot be explained within the ``ideal'' flow
picture. It is why below we propose the hydrodynamics scenario of
the associated mass $M^{\left( \mathrm{ass}\right)  }$ formation
that takes into account the viscous part of this problem. The
straightforward accounting for the non-zero viscosity results in a
dramatic growth (compared to the ideal case) of kinetic energy of
the moving sphere and, consequently, its effective associated
mass. This fact is caused by setting in motion of a spacious
domain of viscous liquid around the moving particle. Nevertheless,
the arising divergency can be cut off by accounting for the
non-linear effects in stationary flow.

Another way to introduce the effective associated mass for the
problem under consideration is based on the analysis of the real
experiments where the \textbf{dynamical parameters} of charged
carriers are always measured in oscillating regime. In this
approach the problem can still be assumed linear but the effective
mass becomes frequency dependent. The proposed consideration
permits us to explain consistently the specific features of the
observed on experiments snowball effective mass temperature
dependence in wide interval of temperatures.

Let us start from the definition of the effective associated mass
within the "snow-cloud" approximation and using the two-fluid
model. In that case

\begin{equation}
M^{\left(  \mathrm{ass}\right)  }=M_{s}^{\left(  \mathrm{ass}\right)  }%
+M_{n}^{\left(  \mathrm{ass}\right)  }.\label{ass}%
\end{equation}
The first term can be defined as follows:%

\begin{equation}
M_{s}^{\left(  \mathrm{ass}\right)  }V^{2}/2=\frac{\rho_{s}\left(  T\right)
}{2}\int_{R_{\ast}}^{\infty}\mathbf{v}_{s}^{2}\left(  \mathbf{r}\right)
d^{3}\mathbf{r}, \label{ass1s}%
\end{equation}
where $\rho_{s}\left(  T\right)  $ is the super-fluid density, $V$\ is the
velocity of a "snow-cloud" center of mass forward motion, $\mathbf{v}%
_{s}\left(  \mathbf{r}\right)  $ is the super-fluid component
local velocity distributions appearing in liquid due to charge
carrier motion. Let us underline that some auxiliary parameter
$R_{\ast}$ instead of the snowball radius $R_{s}^{\rm He}$ (see
Eq.(\ref{atkrad})) here is used as the lower limit cut-off. Its
origin and evaluation will be discussed below (see
Eq.(\ref{rid})).

The super-fluid flow has a potential character \cite{Lamb,LL}%

\begin{equation}
\mathbf{v}_{s}=\mathbf{\nabla}\phi\left(  \mathbf{r}\right)  ,\quad\phi\left(
\mathbf{r}\right)  =-\frac{\mathbf{A\cdot n}}{\mathbf{r}^{2}},\quad
\mathbf{A=V}R_{\ast}^{3}/2. \label{supervel}%
\end{equation}
Correspondingly the super-fluid part of \ associated mass $M_{s}^{\left(
\mathrm{ass}\right)  }$ (\ref{ass1s}) with the velocity distribution
determined by the Eq. (\ref{supervel}) is reduced to the expression
(\ref{massb}) with the simple substitutions $R_{s}\rightarrow R_{\ast}$ and
$\rho\rightarrow\rho_{s}$.

As to the normal fluid part of the associated mass $M_{n}^{\left(
\mathrm{ass}\right)  }$, its definition turns out to be more
cumbersome. It would be natural to define it in the same way as
the superfluid one, just substituting in \ Eq. (\ref{ass1s}) the
subscript $s$\ by $n$. But this programme runs against the
considerable obstacles in spite of the fact that the problem of
rigid sphere motion in a viscous liquid was considered a long ago
(see, for instance, \cite{Lamb,LL}). Indeed, for small Reynolds
numbers $\operatorname{Re}=\rho VR_{\ast}/\eta\ll1$ and distances
$R_{\ast}<r<R_{\ast }/\operatorname{Re}$ the gradient term $\left(
\mathbf{v}\cdot\mathbf{\nabla }\right)  \mathbf{v}$ in
corresponding Navier-Stokes equation can be ignored. The solution
of such linearized equation in zero approximation turns out to be
independent on the liquid viscosity $\eta$. In the center of mass
frame it has the form:
\begin{align}
v_{r}\left(  r,\theta\right)   &  =-V\cos\theta\left(  \frac{3R_{\ast}}%
{r}-\frac{R_{\ast}^{3}}{2r^{3}}\right)  ,\label{ur2}\\
v_{\theta}\left(  r,\theta\right)   &  =V\sin\theta\left(  \frac{3R_{\ast}%
}{4r}+\frac{R_{\ast}^{3}}{4r^{3}}\right)  .\label{uta}%
\end{align}
Here $\theta$ is the polar angle counted from the x-axes, which coincides with
the ion velocity direction.

One can easily see, that corresponding contribution to the kinetic
energy diverges at the upper limit of integration. To regularize
this divergence it is necessary to use at large distances
$r\gtrsim l_{\eta}=R_{\ast }/\operatorname{Re}$
($l_{\eta}=\eta/\rho V$ is the characteristic viscous length) more
precise, so-called Ossen, solution of the Navier-Stokes equation
which was obtained taking into account the gradient term in it
\cite{LL}. \ This solution shows that almost for all angles,
besides the domain restricted by the narrow paraboloid
$\theta\left(  x\right)  =\pi-\sqrt {\eta/\rho V|x|}$ behind the
snowball (so-called "laminar trace") at large distances the
velocity field decays exponentially. In latter the velocity decays
by power law \cite{LL} and it gives logarithmically large
contribution with respect to other domain of disturbed viscous
liquid. Such specification permits to cut off formal divergency of
the kinetic energy and to find with the logarithmic accuracy the
value of the normal component of associated mass for the
stationary moving in viscous liquid charge carrier:
\begin{align}
M_{n\left(  st\right)  }^{\left(  \mathrm{ass}\right)  } &  \sim
M_{0}^{\left(  \mathrm{ass}\right)  }\left(  \rho_{n},R_{\ast}\right)  \left(
\frac{l_{\eta}}{R_{\ast}}\right)  \ln\frac{l_{\eta}}{R_{\ast}}\nonumber\\
&  =R_{\ast}^{2}\frac{\eta}{V}\ln\frac{\eta}{\rho VR_{\ast}}.\label{mst1}%
\end{align}
Here the associated mass $M_{0}^{\left(  \mathrm{ass}\right)  }\left(
\rho_{n},R_{\ast}\right)  \ $is defined by Eq.(\ref{massb}) with radius
$R_{\ast}$ and density $\rho_{n}$. Since in our assumptions $l_{\eta}\gg
R_{\ast}$ the value of $M_{n\left(  st\right)  }^{\left(  \mathrm{ass}\right)
}$ turns out much larger than the value of the associated mass in the ideal
liquid. Moreover, when velocity $V\rightarrow0$ such definition,\ in spite of
the performed above regularization procedure, fails since Eq.(\ref{mst1}) diverges.

At this point it is necessary to recall that, as it was mentioned
in introduction, in reality it is the dynamic associated mass
which has to be considered. This circumstance allows one to
incorporate qualitatively new features in the effective mass
behavior. Indeed, let us consider the situation, when a periodic
electric field is applied to a charge carrier placed in liquid
He$^{4}$. The dynamic Stokes force, appearing when the sphere
oscillates in the viscous liquid with finite
frequency, has the form \cite{Mellor2,Lamb}:%

\begin{align}
F\left(  \omega\right)   &  =6\pi\eta R_{\ast}\left(  1+\frac{R_{\ast}}%
{\delta\left(  \omega\right)  }\right)  V\left(  \omega\right)  \label{fomega}%
\\
&  +3\pi R_{\ast}^{2}\sqrt{\frac{2\eta\rho_{\mathrm{n}}}{\omega}}\left(
1+\frac{2R_{\ast}}{9\delta\left(  \omega\right)  }\right)  i\omega V\left(
\omega\right)  ,\nonumber
\end{align}
where $\delta\left(  \omega\right)  =\left(  2\eta/\rho_{n}\omega\right)
^{1/2}$ is so-called dynamic penetration depth. It is natural to identify the
coefficient in front of the Fourier transform of acceleration $\ i\omega
V\left(  \omega\right)  $\ with the effective dynamic associated mass, what
gives:%
\begin{equation}
M_{n \left(  dyn\right) }^{\left(  \mathrm{ass}\right)  }\left(
\omega\right)  =M_{0}^{\left(  \mathrm{ass}\right)  }\left(
\rho_{n},R_{\ast
}\right)  \left(  1+\frac{9}{2}\frac{\delta\left(  \omega\right)  }{R_{\ast}%
}\right)  .\label{momega}%
\end{equation}
One can see that for high frequencies ($\delta\left(
\omega\right)  \ll R_{\ast}$) the dynamic associated mass
coincides with that one of a sphere moving stationary in an ideal
liquid, while when $\omega\rightarrow0$ $M_{n\left(  dyn\right)
 }^{\left(  \mathrm{ass}\right)  }\left(
\omega\right)  $ diverges as $\omega^{-1/2}$. As we already have seen above
this formal divergence is related to fall down of the linear approximation in
the Navier-Stokes equation assumed in derivation of Eq.(\ref{fomega}). \ It is
clear that the definition (\ref{momega}) is valid for high enough frequencies,
until $\delta\left(  \omega\right)  \lesssim l_{\eta}$ (i.e. $\omega
\gtrsim\widetilde{\omega}=\eta/\rho_{n}R_{\ast}^{2}).$When $\omega$ becomes
lower than $\widetilde{\omega}$ the penetration depth $\delta\left(
\omega\right)  $ in Eq.(\ref{momega}) has to be substituted by $l_{\eta}$ and
up to the accuracy of $\ln l_{\eta}/R_{\ast}$ the dynamic definition
Eq.(\ref{momega}) matches with the static one (see Eq.(\ref{mst1})).

Thus, the following qualitative picture arises. In the stationary
regime, the moving sphere disturbs an enormous domain in viscous
liquid. This perturbation results in the large effective mass
(\ref{mst1}) and can substantially affect the critical conditions
for the generation of vertex rings. When going to the oscillatory
regime the viscous "coat" and especially the "laminar trace"  do
not keep pace with the charge in its oscillations, the perturbed
domain shrinks with the growing frequency thus reducing the value
of effective mass.

The question arises, how sensitive are the introduced above
definitions of the associated mass Eqs.(\ref{momega}),
(\ref{ass1s}) to the real shape of the liquid density perturbation
in the vicinity of the charge carrier? The examples are already
known when the snowball Atkins's model and more realistic
snow-cloud model lead to qualitatively different predictions for
the value of positive ion mobility in liquid He$^{4}$ (see
\cite{LL}-\cite{Bowley} ). In order to clear up this problem let
us consider the hydrodynamic picture of a snow-cloud motion. The
latter we assume as the density compression which
decays with distance by power law:%

\begin{equation}
\rho\left(  r\right)  =\rho+\delta\rho\left(  r\right)  =\rho+\frac{C}{r^{4}%
}\label{powden}%
\end{equation}
(\ constant $C$ is expressed in terms of Eqs.(\ref{chempot})). The continuity
equation is read as%

\begin{equation}
\rho\left(  r\right)  div\mathbf{v+v\cdot\nabla}\rho\left(  r\right)  =0.
\label{coneq}%
\end{equation}
For an ideal (superfluid) liquid the velocity field is potential and it can be
presented in the form Eq.(\ref{supervel}). Substituting Eqs.(\ref{powden}) and
(\ref{supervel}) to Eq.(\ref{coneq}) one finds%

\begin{equation}
\Delta\phi\left(  \mathbf{r}\right)  -\frac{4C}{r^{5}}\frac{\mathbf{\nabla
}\phi\left(  \mathbf{r}\right)  }{\rho+\delta\rho}=0. \label{poisson1}%
\end{equation}
The Eq.(\ref{poisson1}) has to be solved with the additional requirement%

\begin{equation}
\phi\left(  r\rightarrow\infty\right)  \rightarrow rV\cos\theta.
\end{equation}

In Born approximation the Eq.(\ref{poisson1}) is reduced to the
Poisson equation. Supposing that $\phi\left(  r\right)
=rV\cos\theta+\phi_{1}\left(  r\right)
,$ it can be rewritten in the form%

\begin{equation}
\Delta\phi_{1}-\frac{4C}{r^{5}}\frac{V\cos\theta}{\rho+\delta\rho}%
=0,\qquad\phi_{1}\left(  r\rightarrow\infty\right)  \rightarrow0,
\end{equation}
which solution can be presented as

\begin{equation}
\phi_{1}\left(  \mathbf{x}\right)  =\frac{\mathbf{d\cdot x}}{r^{3}}%
,\qquad\mathbf{d}=\frac{C}{\pi}\int\frac{\mathbf{r}}{r^{6}}\frac{\left(
\mathbf{V\cdot r}\right)  }{\rho+\delta\rho\left(  r\right)  }d^{3}\mathbf{r.}
\label{fi1sol}%
\end{equation}

The Eq.(\ref{fi1sol}) shows, that far enough from a snow-cloud center the
potential of velocity distribution looks like a dipole one, exactly in the
same way as for the flow around the rigid sphere of the radius
$R_{\mathrm{ideal}}$. Using Eqs.(\ref{supervel}) and (\ref{fi1sol}) one finds%

\begin{equation}
R_{\mathrm{ideal}}^{3}=2\pi C\int_{0}^{\infty}\frac{dr}{r^{2}\left(
\rho+\delta\rho\right)  }. \label{rid}%
\end{equation}
It is easy to see that in spite of presence of $r^{2}$ in the
denominator this integral converges at the lower limit.

The above analysis demonstrates that the problem of the associated mass
definition in the snowball and the snow-cloud models for an ideal liquid are
qualitatively identical. It is just enough to renormalize the effective radius
$R_{\ast}$ of the hydrodynamic dipole. Elsewhere will be shown that the same
conclusion can be done basing on the Navier-Stokes formalism for the motion of
the snow-cloud in the viscous liquid (naturally, with other definition of the
effective radius).

Let us pass to discussion of the experimental situation. First of
all it is necessary to stress that the effective mass temperature
dependence for positive charge carriers in liquid He$^{4}$ has
been observed in the interval from $50\,mK$ up to $2\,K$ (see
\cite{Mellor1}-\cite{Mellor2} ). The growth of the mass with
temperature occurs with the average rate $28\,m_{{\rm He}^{4}}/K.$
The viscous scenario does not permit us to analyze the region of
very low temperatures, so let us restrict our consideration by the
interval $0.5\,K\leq T\leq2\,K,$ where the density of the normal
component varies in the limits $0.005\,\leq\rho_{n}/\rho\leq0.7.$
The data concerning temperature dependence $M^{\left(
\mathrm{ass}\right)  }\left(  T\right)  $ were taken from the
\textit{ac }transport measurements. Namely, the authors of Ref.
\cite{Dahm} analyzed temperature dependence of the position of
maximum observed in frequency dependent conductivity. This
position in our model evidently corresponds to situation when both
the real and imaginary parts of the Stokes force (\ref{fomega})
are of the same order. Equating them one can evaluate the
corresponding frequency as%

\begin{equation}
\omega_{x}\left(  T\right)  =\frac{x^2\eta\left(  T\right)
}{2\rho_{n}\left( T\right)  R_{\ast}^{2}}\qquad x^2=9/2.
\end{equation}
Corresponding value of the associated mass is
\begin{align}
M_{n \left(  dyn\right) }^{\left(  \mathrm{ass}\right)  }\left(
\omega_x\right)= 3\pi
R_*^3\rho_n\frac{1}{x}\left(1+\frac{2}{9}x\right).\label{masfin}
\end{align}

\begin{figure}[h]
\rotatebox{270}{\includegraphics*[width=5.0cm]{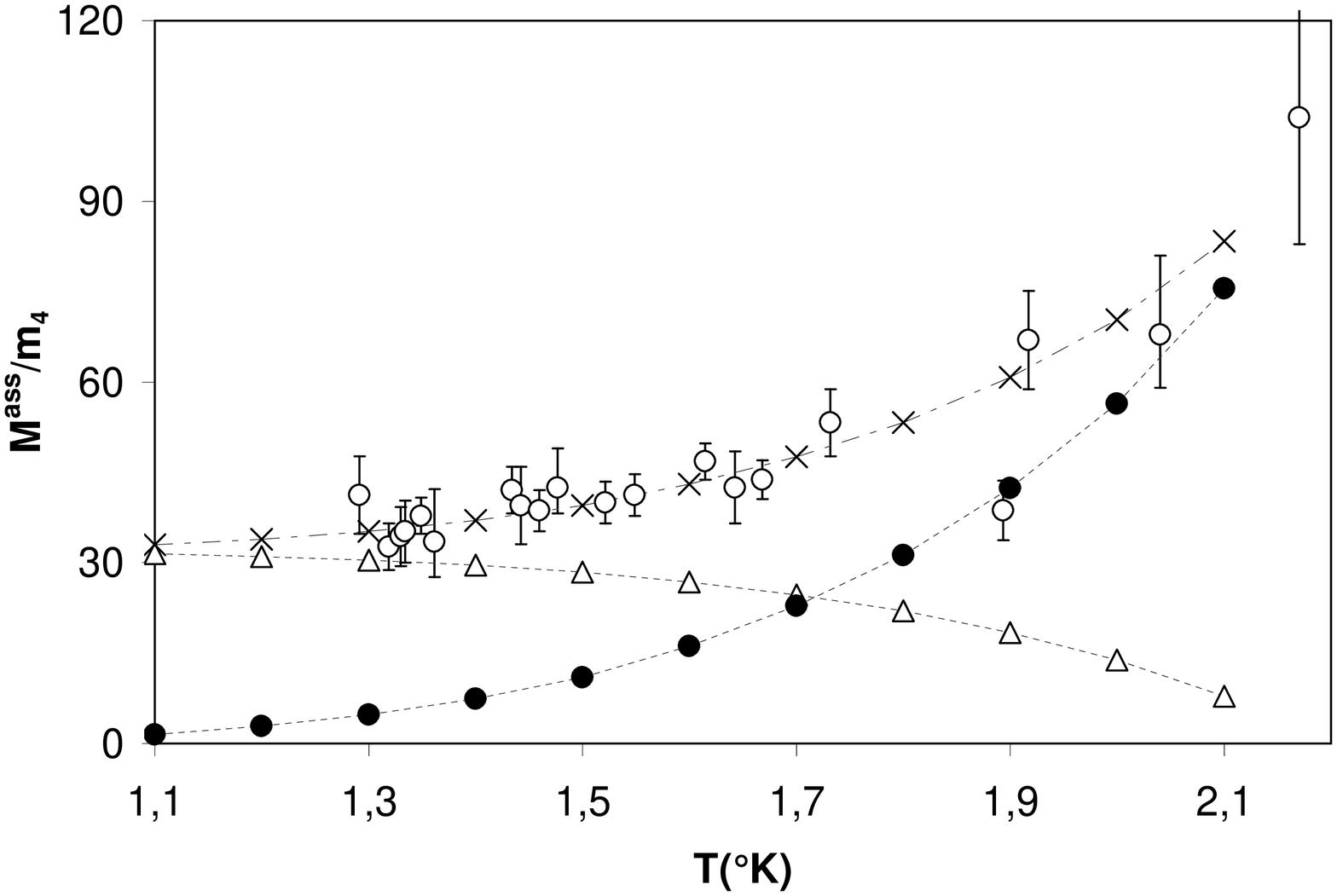}}
\caption{ The temperature dependencies of $M_{s}^{\left(
\mathrm{ass}\right) }\left( T\right)$ according to Eqs.
(\ref{ass1s}, \ref{supervel}) (triangles), $M_{n}^{\left(
\mathrm{ass}\right)  }\left( T\right)$ (see Eq. (\ref{masfin}),
black dots) and their sum $M_{s}^{\left( \mathrm{ass}\right)
}\left( T\right) +$ $M_{n}^{\left( \mathrm{ass}\right)  }\left(
T\right) $ (crosses) are presented separately. The dependence
$\rho_{n}\left( T\right) $ is taken
from \cite{Putter}. The experimental data (open circles with error bars) are taken from Ref. \cite{Dahm}}%
\label{assmass}%
\end{figure}

Fitting of the Eq. (\ref{ass})\  with the Eqs. (\ref{ass1s},
\ref{supervel}) for $M_{s}^{\left( \mathrm{ass}\right) }$ and
(\ref{masfin}) for $M_{n }^{\left( \mathrm{ass}\right) }$
demonstrates a good agreement with the experimentally observed
temperature dependence in the wide range of temperatures (see Fig.
\ref{assmass}). This fact strongly supports the idea to use
two-liquid scenario and to account for the viscous velocity
distribution in the frameworks of the Navier-Stokes flow picture.
It is necessary to mention that understanding of possible viscous
origin of the temperature dependence $M_{n}^{\left(
\mathrm{ass}\right)  }\left( T\right)  $ was already contained in
the early paper \cite{Dahm}. Without derivation, just referring to
\cite{LL,Stokes}, the authors presented their definition of the
associated mass, which resembles much our formula (\ref{momega}),
but contained the additional factor $\rho_{n}/\rho$ in its second
term. This difference turns out essential for the fitting of
experimental data, leading to the visible discrepancy at low
temperatures. That is perhaps the reason why in Ref. \cite{Dahm}
the real $M(T)$ fitting is only realized in a narrow range $\pm
0.1 K$ in the vicinity of $2.1 K $, where $\rho_n/\rho \to 1$.

In conclusion, it should be emphasized that the observed
temperature (actually, frequency) dependence of the snowball
effective mass is interesting not only in itself. It also turns
out sensitive indicator revealing qualitative difference in the
velocity distribution field around the moving sphere in either
viscous or ideal regime. This difference has been known for a long
time. However, it did not attract much attention since slow
decrease of the velocity field ($\sim 1/r$) does not result in any
divergence in the real part of the Stokes drag force. The
situation is quite different in calculation of its imaginary part,
and outlined above insight is one of important conclusions of this
paper.

Certainly, the hydrodynamic treatment of the $M(T)$ dependence
imposes some restrictions on explanation of the available
experimental data obtained at low temperatures (in the vicinity of
several mK). However, the alternative kinetic language suitable
for description of the ballistic regime confronts in substantial
difficulties when the sphere effective mass is calculated even in
the laminar limit. Details of this formalism will be reported
elsewhere.

This work was partly supported by RFBR grant No 06 02 17121 and
the Program of the Presidium of Russian Academy of Sciences
``Physics of Condensed Matter''.

\end{document}